**Meingast et al. reply:** In this Reply we address a Comment by R.S. Markiewicz [1] on our Letter [2].

In classical superconductors Cooper-pair formation and phase coherence occur simultaneously as the temperature is lowered below $T_c$. In high-temperature superconductors (HTSC), on the other hand, the small superfluid density and low associated phase stiffness of the superconducting condensate are expected to lead to a separation of the Cooper-pair formation and the phase-coherence temperatures, especially in underdoped materials [3]. The only real phase transition in this scenario is the 3d-XY phase-ordering transition at $T_c$ [3,4]. In our Letter [2] we showed that $T_c$ in underdoped and optimally doped YBCO is just such a phase-ordering temperature, and then the question naturally arises - where do the Cooper pairs form? The observed 3d-XY scaling of our thermal expansion data over a wide temperature range [2] suggests that pairing occurs at temperatures considerably above $T_c$, and it thus appeared quite natural for us to associate the opening of the pseudogap at $T^*$ with Cooper-pair formation [2]. We did not provide a temperature scale for $T^*$, because we do not directly observe it in our data. Our $T_c^{MF}$, which quantifies the strength of the fluctuations, should not be confused with $T^*$.

Markiewicz [1] criticizes the above scenario because he *incorrectly* assumes that superconducting fluctuations should extend all the way up to $T^*$. The mere presence of a pairing amplitude is, however, not sufficient for superconducting fluctuations, because these also require some phase coherence. *In fact it is the phase-coherence temperature, i.e. $T_c$, and not $T^*$, which sets the temperature scale for superconducting fluctuations.* For example, in superfluid $^4$He there are no signs of superfluidity considerably above $T_\lambda$ even though there exists a full amplitude of 'preformed pairs' (bosons) up to much higher temperatures. The onset of phase fluctuations ($T_{fluct.}^{onset}$) for YBCO as a function of doping (see Fig. 1) can be estimated using the simple anisotropic 3d-XY model, in which $T_{fluct.}^{onset}$ is roughly 1.5 x $T_c$ and 2 x $T_c$ for the isotropic (3d) and nearly 2-dimensional (2d) cases, respectively (see Fig. 2b of Ref. [2]). In Fig. 1 we have assumed that YBCO is in the 2d limit in the strongly underdoped region ($T_{fluct.}^{onset}$ =2 $T_c$), then smoothly starts to cross over to 3d behavior somewhat below optimal doping, becomes 3d at optimal doping ($T_{fluct.}^{onset}$ =1.5 $T_c$), and finally crosses over to mean-field like behavior in the overdoped region ($T_{fluct.}^{onset}$ = $T_c$). Fig.1 shows that $T_{fluct.}^{onset}$ in the 3d-XY approach essentially follows $T_c$ and not $T^*$, as claimed in Ref. [1]. $T_c^{MF}$, as determined in our

Letter [2], should always lie between $T_c$ and $T_{\text{fluct.}}^{\text{onset}}$. In the 3d-XY approach, the phase transition is due to the unbinding of vortex loops [4], and one expects vortex-like correlations to persist above $T_c$ in the phase-fluctuation region bounded by $T_{\text{fluct.}}^{\text{onset}}$. Such correlations have been observed using the Nernst effect [5], and the doping dependence of the onset of this vortex-like Nernst signal is very similar to our $T_{\text{fluct.}}^{\text{onset}}$, providing further support for 3d-XY scaling in underdoped HTSC. Finally, recent theoretical works [6] indicate that much of the anomalous 'normal-state' properties of HTSC can be understood in the framework of this precursor superconductivity scenario.

C. Meingast,[1] V. Pasler,[1] P. Nagel,[1] A. Rykov,[2] S. Tajima,[2] and P. Olsson[3]

[1]Forschungszentrum Karlsruhe, Institut für Festkörperphysik, 76021 Karlsruhe, Germany

[2]Superconductivity Research Lab-ISTEC, 10-13 Shinonome 1-chome, Koto-ku, Tokyo, Japan

[3]Department of Theoretical Physics, Umeå University, S90187 Umeå, Sweden

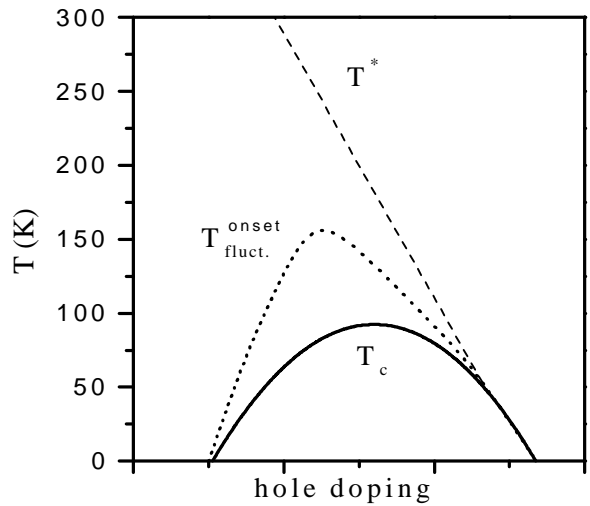

Fig. 1. Onset of phase fluctuations, $T_{onset}^{fluct.}$, $T_c$ and the pairing (or pseudogap) temperature $T^*$ for YBCO within a simple 3d-XY model. (see text for details)